\documentclass[12pt]{iopart}
\usepackage{graphicx}
\usepackage{amsfonts}
\usepackage{amssymb}

\begin{document}
\title[Cosmological Alfv\'en waves and the observational bound on the vector perturbation]{Cosmological Alfv\'en waves in the recent CMB data, and the observational bound on the primordial vector perturbation}
\author{Jaiseung Kim$^1$ and Pavel Naselsky$^1$}
\address{$^1$ Niels Bohr Institute, Blegdamsvej 17, DK-2100 Copenhagen, Denmark}
\ead{jkim@nbi.dk}

\begin{abstract}
In the presence of the primordial magnetic field, initial vector (vorticity) perturbations produce cosmological Alfv\'en waves and leave imprints on cosmic microwave background (CMB) temperature and polarization anisotropy.

We have investigated imprints of cosmological Alfv\'en waves in CMB anisotropy. 
For data constraints, we have used the power spectrum of the recent CMB observations, and correlations estimated from WMAP Internal Linear Combination (ILC) maps. 
Our analysis shows 3$\sigma$ evidence of cosmological Alfv\'en waves.
Using the 3$\sigma$ limit from our analysis and the Alfv\'en velocity limit from the total energy density constraint, we impose a lower bound on the amplitude of primordial vector perturbation: $4\times 10^{-12}$  at $k_0=0.002/\mathrm{Mpc}$. 
\end{abstract}


\section{Introduction}
There are strong observational evidences on the existence of large-scale magnetic fields, whose strength is on the order of micro Gauss \cite{Magnetic_Fields_Lyman,Magnetic_Fields_Galaxy,Magnetic_Fields_Origin,Magnetic_Fields_Faraday}. 
These magnetic fields are believed to be seeded by a small Primordial Magnetic Field (PMF) \cite{PMF_Cosmic_transition,PMF_Electroweak,PMF_QCD,PMF_quark}. 
Primordial Magnetic Field may be considered to consist of a homogeneous and inhomogeneous (stochastic) magnetic fields.
In this paper, we focus on the effect of Primordial Magnetic Field on CMB anisotropy associated with vector perturbations.
Primordial vector perturbation, whose source includes topological defects and inhomogeneous primordial magnetic fields \cite{cosmic_defect_Vector,cosmic_defect_CMB,SPMF_CMB}, decays rapidly with the expansion of the Universe, and therefore, does not leave observable signature on CMB anisotropy \cite{Foundations_Cosmology,Cosmology}.
However, primordial vector perturbation in the presence of a homogeneous primordial magnetic field induces Alfv\'en waves in the cosmic plasma, which may leave observable imprints on CMB anisotropy via a Doppler effect and an integrated Sachs-Wolfe effect \cite{PMF_acoustic_CMB,inhomogeneous_Alfven,pmf_small,pmf_pol,DKY,Alfven_nl}. 
To be specific, cosmological Alfv\'en waves create correlations between $a_{lm}$, $a_{l\pm\Delta l,m\pm\Delta m}$, where $\Delta l=0, 1, 2$ and $\Delta m=0,1,2$ \cite{KLR}. Over the past several years, there have been great success in measurement of Cosmic Microwave Background (CMB) anisotropy by ground and satellite observations \cite{WMAP5:basic_result,WMAP5:powerspectra,WMAP5:parameter,ACBAR,ACBAR2008,QUaD1,QUaD2,QUaD:instrument}, and observational imprints of cosmological Alfv\'en waves in CMB anisotropy have been studied by several authors \cite{DKY,KLR,Chen,Naselsky:PMF,PMF_anomaly}. By investigating the effect on power spectrum and signature correlations in the recent CMB data, we attempt to constrain cosmological Alfv\'en waves, and also impose a lower bound on the primordial vector perturbation.

The outline of this paper is as follows. 
In Section \ref{cmb_pmf}, we briefly review the effect of cosmological Alfv\'en waves on CMB anisotropy.
In Section \ref{correlation}, we discuss statistical properties of CMB in the presence of the Alfv\'en waves.  
In Section \ref{estimators}, we build the statistics sensitive to the imprints of Alfv\'en waves. 
In Section \ref{analysis}, we analyze the recent CMB data and present the result.
In Section \ref{Discussion}, we make summary and conclusion.
In \ref{leakage}, we discuss the effect of incomplete sky coverage on the analysis of Alfv\'en wave imprints.

\section{The effect of Alfv\'en waves on CMB anisotropy}
\label{cmb_pmf}
According to linearized Einstein equations \cite{Spacetime}, vector perturbation decays rapidly with the expansion of the Universe, and therefore, does not leave observable signature on CMB anisotropy, unless the initial values of vector perturbation are unusually high \cite{Foundations_Cosmology,Cosmology}.
In the presence of a homogeneous magnetic field, vector perturbation induces Alfv\'en waves in cosmic plasma, and leaves observable imprints on CMB anisotropy via a Doppler effect and an integrated Sachs-Wolfe effect \cite{DKY}. Durrer, Kahniashvili and Yates \cite{DKY} (hereafter, DKY) showed that cosmological Alfv\'en waves generate the fractional CMB anisotropy for a Fourier mode $\mathbf{k}$:
\begin{eqnarray}
\frac{\delta T}{T_0}(\hat {\mathbf{n}},\mathbf{k}) \approx \hat {\mathbf{n}}\cdot \mathbf{\Omega}(\mathbf k,\eta_{dec})
= \hat {\mathbf{n}}\cdot \mathbf{\Omega_0}\; v_A k\,\eta_{dec}\;\hat {\mathbf{B}}\cdot \hat {\mathbf{k}}, 
\end{eqnarray}
where $\hat {\mathbf{n}}$ and $\hat {\mathbf{B}}$ denote sky direction and a homogeneous magnetic field direction respectively, $\mathbf{\Omega}(\mathbf k,\eta_{dec})$ is the Gauge invariant linear combination associated with vector perturbations, $\eta_{dec}$ denotes the conformal time at the moment of baryon-photon decoupling, $v_A$ is Alfv\'en wave velocity and $T_0$ is the CMB monopole temperature $2.725 [\mathrm{K}]$ \cite{Fixen:dipole}. 
DKY assumed that vector perturbations are initially created by some random process and have the following statistical properties over an ensemble of universes:
\begin{eqnarray}
\langle \Omega^i_0(\mathbf k)\,\Omega^j_0(\mathbf k)\rangle =(\delta_{ij}-\hat k_i\,\hat k_j) P(k),
\end{eqnarray}
$P(k)$, which is the power spectrum, is assumed to follow a simple power law:
\begin{eqnarray} 
P(k)=A_v \frac{k^n}{k^{n+3}_0},\label{P} 
\end{eqnarray}
where $k_0$ is a pivot wavenumber and we set it to $0.002/\mathrm{Mpc}$, which is equal to the 
WMAP team's pivot wavenumber for scalar perturbation \cite{WMAP3:parameter}. 
Possible sources of primordial vector perturbation include inhomogeneous stochastic primordial magnetic fields and topological defects \cite{cosmic_defect_Vector,cosmic_defect_CMB,SPMF_CMB}.

\section{CMB statistical properties}
\label{correlation}
The temperature anisotropy $\delta T(\theta,\phi)$ is conveniently decomposed in terms of spherical harmonics $Y_{lm}(\theta,\phi)$ :
\begin{eqnarray}
\delta T(\theta,\phi)=\sum_{lm} a_{lm}\,Y_{lm}(\theta,\phi),
\end{eqnarray}
where $a_{lm}$ are the coefficients of decomposition, and $\theta$ and $\phi$ are a polar and an azimuthal angle.
Kahniashvili, Lavrelashvili and Ratra \cite{KLR} (hereafter KLR) showed that the CMB anisotropy in the presence of cosmological Alfv\'en waves has the following statistical properties:
\begin{eqnarray} 
\lefteqn{\langle a^*_{lm} a_{lm}\rangle = C^{0}_l+(3\cos^2\theta_B-1)\frac{l(l+1)}{(2l-1)(2l+3)}}\nonumber\\
&\times&\;\left\{\frac{l(l+1)+(l^2+l-3)\cos^2\theta_B}{3\cos^2\theta_B-1} - m^2\left[1-\frac{3}{l(l+1)}\right]\right\} I^{l,l}_d\,\label{correlation0} 
\end{eqnarray}
\begin{eqnarray} 
\langle a^*_{lm} a_{l,m+1} \rangle &=& -\sin2\theta_B \exp[-i\phi_B] I^{l,l}_d\,d^{m,m+1}_{ll},\label{correlation1}
\end{eqnarray}
\begin{eqnarray} 
\langle a^*_{lm} a_{l,m+2} \rangle &=& -\frac{1}{2}\sin^2\theta_B \exp[-2i\phi_B] I^{l,l}_d\,d^{m,m+2}_{ll}, \label{correlation2}
\end{eqnarray}
\begin{eqnarray} 
\langle a^*_{l,m} a_{l+2,m} \rangle&=&-(3\cos^2\theta_B-1) I^{l,l+2}_d\,d^{m,m}_{l,l+2}, \label{correlation3}
\end{eqnarray}
\begin{eqnarray} 
\langle a^*_{l,m} a_{l+2,m+1}\rangle&=&\sin2\theta_B\exp[-i\phi_B] I^{l,l+2}_d\,d^{m,m+1}_{l,l+2}, \label{correlation4}
\end{eqnarray}
\begin{eqnarray} 
\langle a^*_{l,m} a_{l+2,m-1}\rangle&=&\sin2\theta_B\exp[i\phi_B] I^{l,l+2}_d\,d^{m,m-1}_{l,l+2}, \label{correlation5}
\end{eqnarray}
\begin{eqnarray} 
\langle a^*_{l,m} a_{l+2,m+2}\rangle&=&-\frac{1}{2}\sin^2\theta_B\exp[-2i\phi_B] I^{l,l+2}_d\,d^{m,m+2}_{l,l+2}, \label{correlation6}
\end{eqnarray}
\begin{eqnarray} 
\langle a^*_{l,m} a_{l+2,m-2}\rangle&=&-\frac{1}{2}\sin^2\theta_B\exp[2i\phi_B] I^{l,l+2}_d\,d^{m,m-2}_{l,l+2},\label{correlation7}
\end{eqnarray}
where $C^{0}_l$ is the power spectrum in the absence of Alfv\'en waves, $\theta_B$ and $\phi_B$ are the spherical coordinate of a PMF direction $\hat {\mathbf B}$, and $d^{m,m'}_{ll'}$ $(|m|\le l,\;|m'|\le l')$ are
\begin{eqnarray*} 
d^{m,m+1}_{l,l} &=&\frac{l^2+l-3}{(2l-1)(2l+3)} (m+\frac{1}{2})\sqrt{(l-m)(l+m+1)},\\
d^{m,m+2}_{l,l} &=& \frac{l^2+l-3}{(2l-1)(2l+3)}\sqrt{(l-m)(l-m-1)(l+m+1)(l+m+2)},\\
d^{m,m}_{l,l+2} &=&\frac{(l+3)l}{2(2l+3)\sqrt{(2l+1)(2l+5)}}\sqrt{((l+1)^2-m^2)(l-m+2)(l+m+2)},\\
d^{m,m+1}_{l,l+2} &=&\frac{(l+3)l}{2(2l+3)\sqrt{(2l+1)(2l+5)}} \sqrt{T((l+1)^2-m^2)(l+m+2)(l+m+3)},\\
d^{m,m-1}_{l,l+2}&=&\frac{(l+3)l}{2(2l+3)\sqrt{(2l+1)(2l+5)}}\sqrt{((l+1)^2-m^2)(l-m+2)(l-m+3)},\\
d^{m,m+2}_{l,l+2} &=& \frac{(l+3)l}{2(2l+3)\sqrt{(2l+1)(2l+5)}}\,\sqrt{\frac{(l+m+4)!}{(l+m)!}},\\
d^{m,m-2}_{l,l+2} &=&\frac{(l+3)l}{2(2l+3)\sqrt{(2l+1)(2l+5)}}\,\sqrt{\frac{(l-m+4)!}{(l-m)!}},
\end{eqnarray*}
and

\begin{eqnarray} 
I^{ll'}_d&=&\frac{2\,T^2_0}{\pi}\int \,k^2 \left(P_0 \frac{k^{n_v}}{k^{n_v+3}_0}\right)\exp\left(-2\frac{k^2}{k^2_D}\right) v^2_A\,\left(\frac{\eta_{dec}}{\eta_0}\right)^2 j_l(k\eta_0)\,j_l'(k\eta_0)\,dk,\nonumber\\
&=&\frac{2\,T^2_0}{\pi}\frac{P_0\,v^2_A}{(k_0\eta_0)^{n_v+3}} \left(\frac{\eta_{dec}}{\eta_0}\right)^2 \int \,x^{n_v+2}\exp\left(-\frac{2 x^2}{(k_D\eta_0)^2}\right)  j_l(x)j_l'(x)\,dx.\nonumber\\\label{Ill}
\end{eqnarray}
$\eta_0$ is the present conformal time, and $v_A$ is the Alfv\'en velocity, which is given by
\begin{eqnarray} 
v^2_A=\frac{B^2_0}{4\pi(\rho_r+p_r)},\;\;\;\;v_A\sim 4\times10^{-4}\frac{B_0}{10^{-9}\mathrm{Gauss}},\label{v_A}
\end{eqnarray}
where $B_0$ is the magnitude of a homogeneous PMF, and
$\rho_r$ and $p_r$ are the density and the pressure of photons \cite{DKY}. $k_D$ denotes the comoving wavenumber of the dissipation scale, due to photon viscosity and given by $\sim 10/\eta_{dec}$ \cite{DKY}. While DKY and KLR have obtained analytic results by neglecting the damping factor $\exp\left(-2k^2/k^2_D\right)$,
we have computed Eq. \ref{Ill} numerically. The damping effect is getting significant on multipoles $l\gtrsim 500$ \cite{DKY}.

Primordial magnetic fields affect the expansion dynamics of the Universe, because of anisotropic pressures associated with primordial magnetic fields \cite{PMF_Constraint}. 
Considering the shear anisotropy due to primordial magnetic fields and the observed magnitude of CMB anisotropy ($\sim 10^{-5}$), Barrow et al. (1997) have derived an upper limit $B_0<2.27\times10^{-9}\,h/75$ Gauss on a homogeneous primordial magnetic field \cite{PMF_Constraint}. Using this upper limit and Eq. \ref{v_A}, DKY showed $B_0\lesssim 10^{-8}\,\mathrm{Gauss}$ and $v_A\lesssim 10^{-3}$ \cite{DKY}.

It is worth to note that correlations are invariant under the parity inversion of a PMF direction $\hat {\mathbf B}$. In other words, $(\theta_B,\phi_B)\rightarrow (\pi-\theta_B,\phi_B+\pi)$) does not affect Eq. \ref{correlation1}, \ref{correlation2}, \ref{correlation3}, \ref{correlation4}, \ref{correlation5}, \ref{correlation6} and \ref{correlation7}.
note that it is the degeneracy of correlation, not that there should be two physical directions.

\section{Estimators}
\label{estimators}
In order to constrain cosmological Alfv\'en waves, we construct the following statistics:
\begin{eqnarray} 
C_l&=&(2l+1)^{-1}\sum_{m} a^*_{lm} a_{lm},\label{D0}\\
D^1_l&=&l^{-1}\sum_{m\ge 0} a^*_{lm} a_{l,m+1},\label{D1}\\
D^2_l&=&(2l-1)^{-1}\sum_{m} a^*_{lm} a_{l,m+2},\label{D2}\\
D^3_l&=&(2l+1)^{-1}\sum_{m} a^*_{l,m}\,a_{l+2,m},\label{D3}\\
D^4_l&=&(2l)^{-1}\sum_{m} a^*_{l,m}\,a_{l+2,m+1},\label{D4}\\
D^5_l&=&(2l-1)^{-1}\sum_{m} a^*_{l,m}\,a_{l+2,m+2}.\label{D5}
\end{eqnarray}
Because of the reality condition $a_{l\,-m}=(-1)^m\,{a_{lm}}^*$, $\sum\limits_{m=-2l} a^*_{lm} a_{l,m+1}$ is always zero. 
Therefore, we defined $D^1_l$ to sum only terms of $m\ge 0$. 
There also exist the following indentities because of the reality condition:
\begin{eqnarray} 
\sum_{m} a^*_{l,m}\,a_{l+2,m+1}=-\left[\sum_{m} a^*_{l,m}\,a_{l+2,m-1}\right]^*,
\end{eqnarray}
\begin{eqnarray} 
\sum_{m} a^*_{l,m}\,a_{l+2,m+2}=\left[\sum_{m} a^*_{l,m}\,a_{l+2,m-2}\right]^*.
\end{eqnarray}
Therefore, $\sum_{m} a^*_{l,m}\,a_{l+2,m-1}$ and $\sum_{m} a^*_{l,m}\,a_{l+2,m-2}$ are redundant and excluded from consideration.
Note that the statistic used in \cite{DKY,Chen} is equivalent to Eq. \ref{D3}.
Using Eq. \ref{correlation0}, \ref{correlation1}, \ref{correlation2}, \ref{correlation3}, \ref{correlation4} and \ref{correlation6},
we may easily show that the expectation values of $C_l$ and $D^i_l$ are
\begin{eqnarray} 
\bar C_l &=& C^0_l+\frac{l(l+1)}{3} I^{l,l}_d, \label{bar_D0}\\
\bar D^1_l &=& -\sin 2\theta_B \exp[-i\phi_B]\,\frac{I^{l,l}_d}{l}\sum_{m\ge 0} d^{m,m+1}_{ll},\label{bar_D1}\\
\bar D^2_l &=& -\frac{1}{2}\sin^2\theta_B \exp[-2i\phi_B]\,\frac{I^{l,l}_d}{2l-1}\sum_{m} d^{m,m+2}_{ll}, \label{bar_D2}\\
\bar D^3_l &=&-(3\cos^2\theta_B-1)\,\frac{I^{l,l+2}_d}{2l+1}\sum_{m} d^{m,m}_{l,l+2},\label{bar_D3}\\
\bar D^4_l &=&\sin 2\theta_B\exp[-i\phi_B]\,\frac{I^{l,l+2}_d}{2l}\sum_{m} d^{m,m+1}_{l,l+2}, \label{bar_D4}\\
\bar D^5_l &=&-\frac{1}{2}\sin^2\theta_B\exp[-2i\phi_B]\,\frac{I^{l,l+2}_d}{2l-1}\sum_{m}d^{m,m+2}_{l,l+2}. \label{bar_D5}
\end{eqnarray}
We may also show that the variance of $D^i_l$ is
\begin{eqnarray} 
\mathrm{Var}(C_l)&\approx& 2/(2l+1)(C^0_l+N_l)^2,\label{Var0}\\
\mathrm{Var}(D^1_l)&\approx& l^{-1}(C^0_l+N_l)^2,\label{Var1}\\
\mathrm{Var}(D^2_l)&\approx& (2l-1)^{-1}(C^0_l+N_l)^2,\label{Var2}\\
\mathrm{Var}(D^3_l)&\approx& (2l+1)^{-1}(C^0_l+N_l)(C^0_{l+2}+N_{l+2}),\label{Var3}\\
\mathrm{Var}(D^4_l)&\approx& (2l)^{-1}(C^0_l+N_l)(C^0_{l+2}+N_{l+2}),\label{Var4}\\
\mathrm{Var}(D^5_l)&\approx& (2l-1)^{-1}(C^0_l+N_l)(C^0_{l+2}+N_{l+2}),\label{Var5}
\end{eqnarray}
where $N_l$ is noise power spectrum, and we have neglected correlations between distinct spherical harmonic modes (i.e. $\langle a^*_{lm} a_{l'm'}\rangle \approx C^0_l\,\delta_{ll'}\delta_{mm'}$).

Though the underlying distribution for primordial vector perturbations is not necessarily Gaussian, the distribution function of $D^i_l-\bar D^i_l$ tends to Gaussian by the central limit theorem \cite{Math_methods}.
Hence, the likelihood function of $\bar D^i_l$, given the data $D^i_l$, is 
\begin{eqnarray}
\mathcal{L}(\bar D^i_l(\lambda_{\alpha})|D^i_l)=\frac{1}{(2\pi)^{\frac{N}{2}}{\left|\mathbf M\right|}^{\frac{1}{2}}}\exp[-\frac{1}{2}(\mathbf D-\mathbf {\bar D}(\lambda_{\alpha}))^{\dagger}\,{\mathbf M}^{-1}\,(\mathbf D-\mathbf {\bar D}(\lambda_{\alpha}))].\nonumber\\\label{likelihood}
\end{eqnarray}
$\mathbf M$ is a $5\,l_\mathrm{max}\times 5\,l_\mathrm{max}$ nearly diagonal matrix, whose diagonal elements correspond to the variances of $D^1_l,\ldots,D^5_l$:
$\mathbf M \approx \mathrm{diag} \left(\mathrm{Var}(D^1_{l_\mathrm{min}}),\,\ldots,\,\mathrm{Var}(D^5_{l_\mathrm{max}})\right)$. 
$N$ is the data vector size (i.e. $5\,l_\mathrm{max}$), and $\lambda_\alpha$ are the parameters associated with Alfv\'en waves : $\lambda_\alpha \in \{A_v v^2_A, n_v, \theta_B,\phi_B\}$. For low $l$, the deviation of $D^i_l-\bar D^i_l$ from Gaussian distribution is getting non-negligible.
However, it does not affect the best-fit values, but only the error bars, since $\mathbf M$ is nearly diagonal, and thereofore the best-fit parameters are, in fact, determined by minimizing $\chi^2$.
Since $\langle (C_l-\bar C_l) \,(D^i_l-\bar D^i_l)\rangle$ is negligible in comparison to the variance of $C_l$ or $D^i_l$, we may construct a full likelihood function as follows:
\begin{eqnarray}
\mathcal{L}(\bar C_l|C_l) \times \mathcal{L}(\bar D^i_l(\lambda_{\alpha})|D^i_l) \label{full_likelihood},
\end{eqnarray}
where $\mathcal{L}(\bar C_l|C_l)$ is the likelihood function associated with CMB power spectra, whose approximate and optimal expression is found in \cite{WMAP1:parameter_method}. 
Note that cosmological Alfv\'en waves affect CMB power spectrum as well, and hence Eq. \ref{bar_D0} should be used for $\bar C_l$ in estimating $\mathcal{L}(\bar C_l|C_l)$. 
Since CMB power spectra and correlations (i.e. $\bar C_l$ and $\bar D^i_l$) possess non-trivial dependence on cosmological parameters (i.e. standard parameters + Alfv\'en wave parameters), we will resort to Markov Chains Montel Carlo (MCMC) likelihood analysis \cite{WMAP1:parameter_method,CosmoMC}.
The discussion on MCMC analysis and dataset used will be give in the next section.

\section{Analysis of the recent CMB observation data}
\label{analysis}
We have investigated the effect of the WMAP team's KQ75 and KQ85 mask on correlations 
(i.e. $D^i_l$ ), and found that significant amount of fictitious correlations are produced by the masks
(refer to \ref{leakage} for details).
Therefore, we find sky maps, which require a Galactic mask, are not suitable for our analysis. Therefore, we have restricted our correlation estimation to whole sky maps obtained by variants of Internal Linear Combination (ILC) method: The WMAP team's ILC map (WILC), Harmonic ILC (HILC) and Needlet ILC (NILC) \cite{WMAP3:temperature,WMAP5:foreground,HILCT,HILCP,NILC}. 
\begin{figure}[htb!]
\centering
\includegraphics[scale=.5]{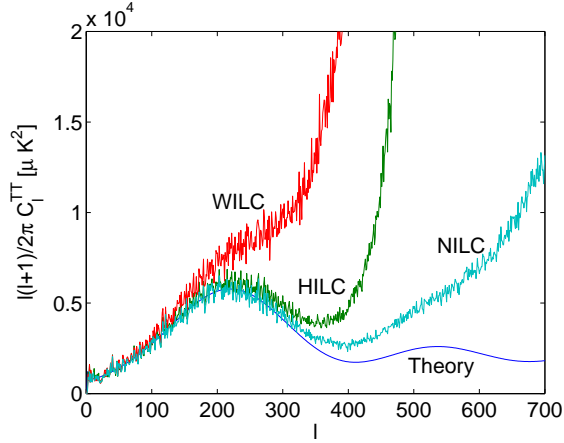}
\caption{the power spectra of ILC maps and the WMAP concordance model}
\label{Cl}
\end{figure}
In Fig. \ref{Cl}, we show the temperature power spectra of the ILC maps with that of the WMAP concordance model \cite{WMAP3:parameter,WMAP5:parameter}. Noting that the power spectra of ILC maps are in good agreement with the model up to $l\le 200$, we have computed $D^i_{l}$ up to $l\le 200$.
We have also made the analysis with $D^i_{l\le 250}$ for HILC, and  $D^i_{l\le 250}$ and $D^i_{l\le 300}$ respectively for NILC.
We have obtained the consistent results with the $D^i_{l\le 200}$ analysis. The confidence intervals are smaller, but the best-fit values turn out to be similar.  From now on, all the results are from the $D^i_{l\le 200}$ analysis.
\begin{table}[htb!]
\centering
\caption{cosmological parameters of $\Lambda$CDM + ($A_v v^2_A$, $n_v$, $\theta_B$, $\phi_B$) constrained by power spectra (WMAP5YR + ACBAR + QUaD) + correlation (ILC) }
\begin{tabular}{cccc}
\hline\hline 
parameter & WILC & HILC & NILC\\
\hline
$\Omega_{b}\,h^2$  &  $0.023^{+0.002}_{-0.002}$ & $0.024^{+0.002}_{-0.003}$ & $0.024^{+0.002}_{-0.003}$\\ 
$\Omega_{c}\,h^2$  &  $0.109^{+0.023}_{-0.024}$ & $0.104^{+0.03}_{-0.018}$ & $0.108^{+0.025}_{-0.024}$\\ 
$\tau$  &  $0.089^{+0.085}_{-0.04}$ & $0.09^{+0.082}_{-0.043}$ & $0.089^{+0.08}_{-0.04}$ \\ 
$n_s$ & $1.015^{+0.262}_{-0.139}$ & $1.028^{+0.242}_{-0.123}$ & $1.018^{+0.251}_{-0.13}$\\ 
${d\,n_s}/{d\ln k}$ &  $-0.008^{+0.068}_{-0.095}$ & $-0.008^{+0.052}_{-0.095}$ & $-0.009^{+0.066}_{-0.108}$\\ 
$\log[10^{10}A_s]$  &  $3.031^{+0.17}_{-0.27}$ & $2.975^{+0.228}_{-0.234}$ & $3.026^{+0.188}_{-0.321}$\\ 
$A_{\mathrm{sz}}$  &  $1.63^{+0.37}_{-1.63}$ & $1.87^{+0.133}_{-1.867}$ & $1.57^{+0.43}_{-1.57}$\\ 
$H_0$  [km/s/Mpc]&  $74.37^{+13.65}_{-10.58}$ & $76.62^{+10.19}_{-13.02}$ & $74.46^{+13.55}_{-10.98}$\\ 
$10^9 A_v\,v^2_A$ &  $3.13^{+5.66}_{-2.42}$ & $4.16^{+4.99}_{-3.27}$ & $3.87^{+5.28}_{-3.11}$\\ 
$n_v$ &  $-4.35^{+0.58}_{-0.7}$ & $-4.45^{+0.6}_{-0.71}$ & $-4.49^{+0.73}_{-0.7}$\\ 
$\theta^+_B$ &  ${15.7^\circ}^{+25}_{-15.7}$ &${1.6^\circ}^{+31.7}_{-1.6}$ & ${11.7^\circ}^{+27.7}_{-11.7}$\\ 
$\phi^+_B$ &  ${112^\circ}^{+247.9}_{-111.8}$ & ${113.2^\circ}^{+246.6}_{-113.1}$ &${119^\circ}^{+240.9}_{-119}$\\ 
$\theta^-_B$ &   ${164.3^\circ}^{+15.7}_{-25}$ & ${178.4^\circ}^{+1.6}_{-31.7}$ & ${168.3^\circ}^{+11.7}_{-27.7}$\\         $\phi^-_B$  &  ${292^\circ}^{+247.9}_{-111.8}$ & ${293.2^\circ}^{+246.6}_{-113.1}$ &${299^\circ}^{+240.9}_{-119}$\\
\hline
\end{tabular}
\label{parameters}
\end{table}

\begin{figure}[htb!]
\centering\includegraphics[scale=.62]{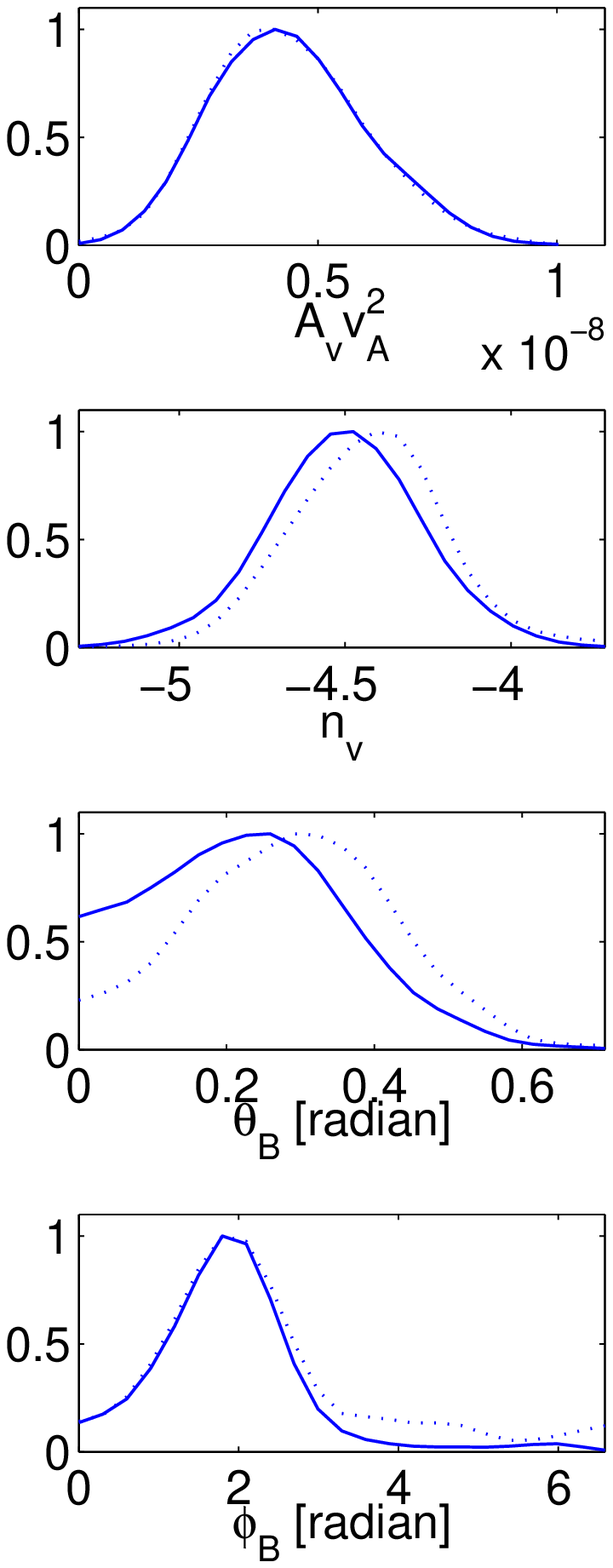}
\centering\includegraphics[scale=.62]{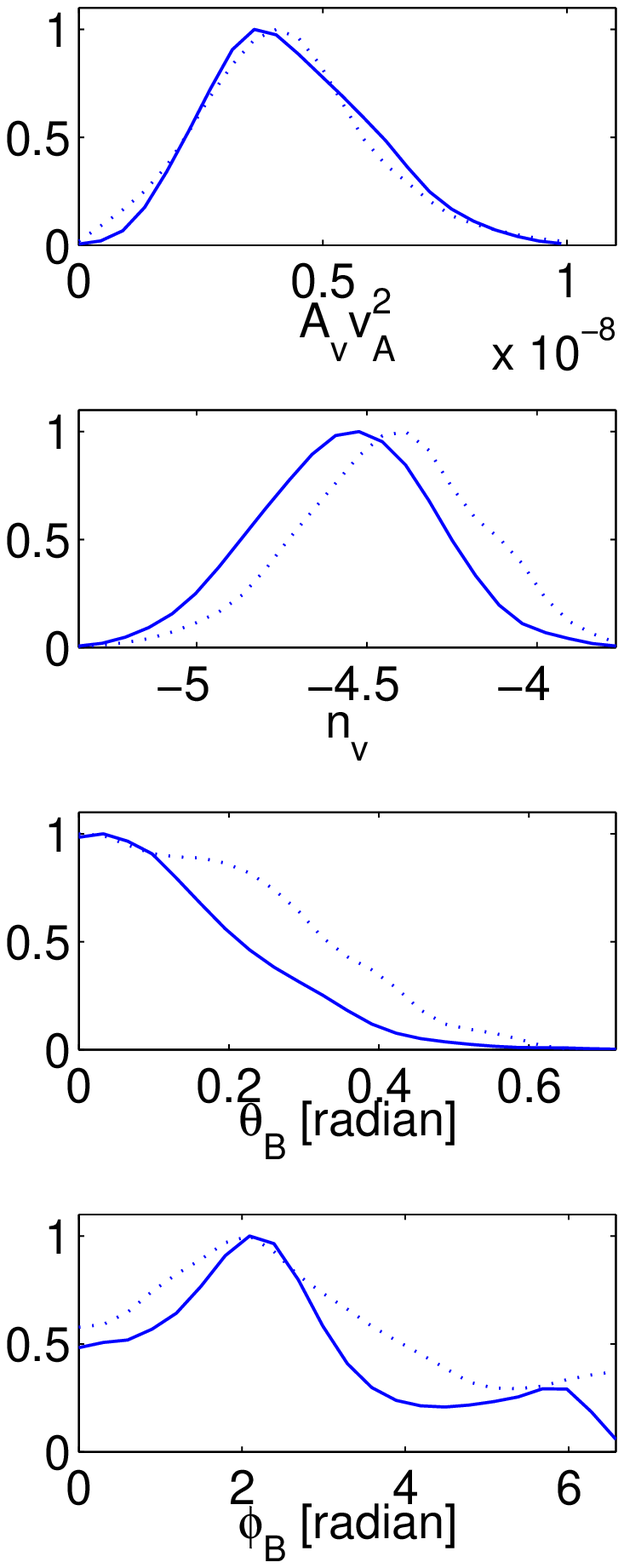}
\centering\includegraphics[scale=.62]{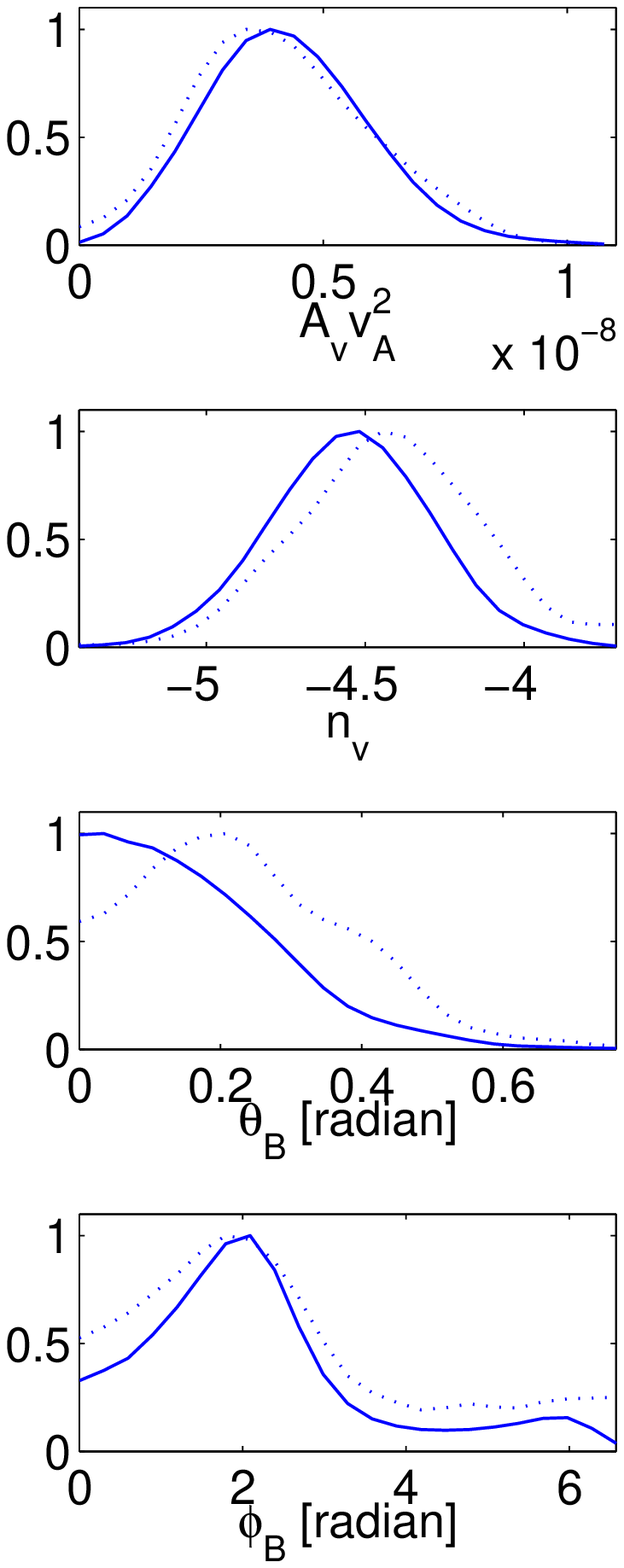}
\caption{Likelihood of Alfv\'en wave parameters: WILC, HILC and NILC (from the left to the right), normalized to its peak.}
\label{likelihood1}
\end{figure}

\begin{figure}[htb!]
\centering\includegraphics[scale=.32]{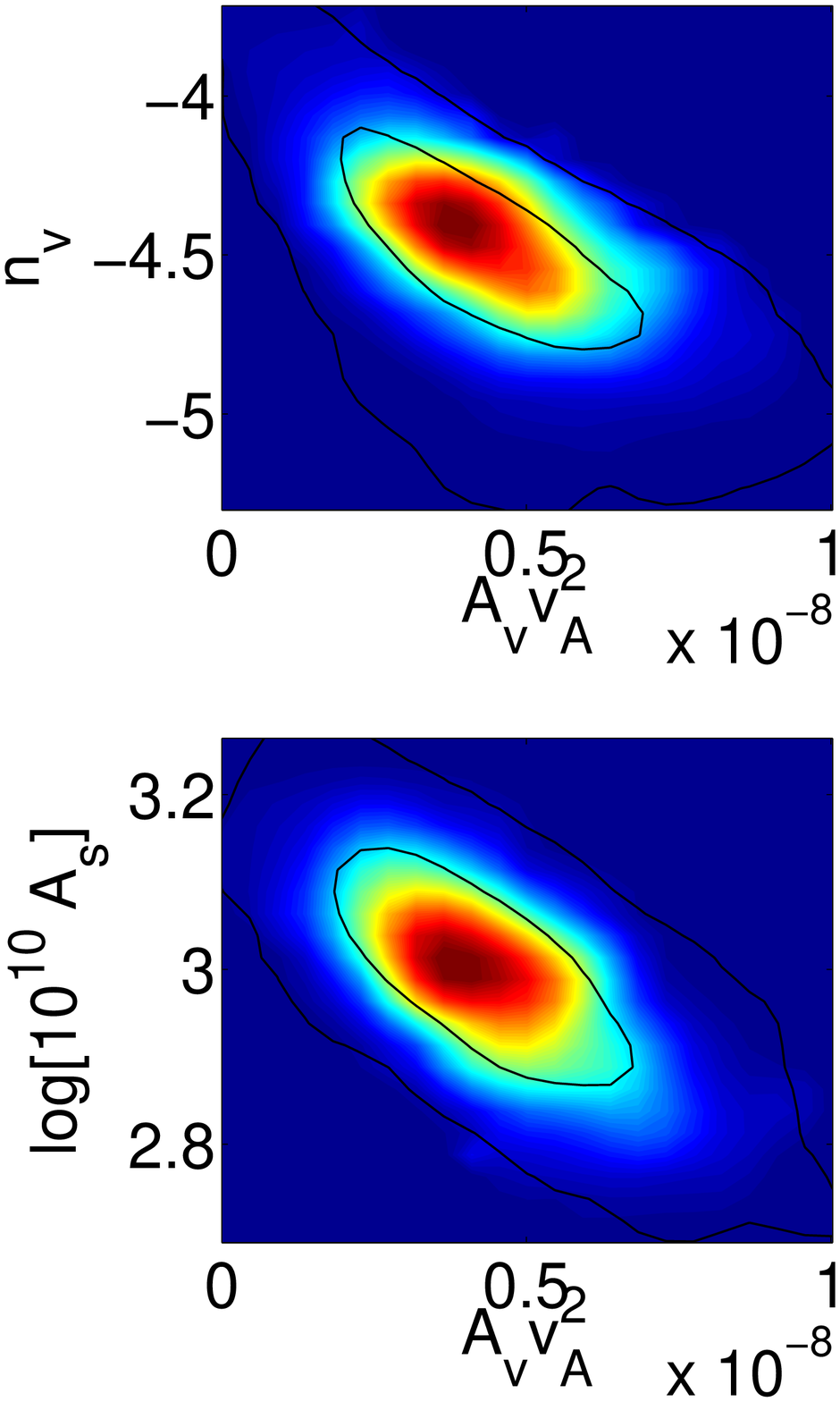}
\centering\includegraphics[scale=.32]{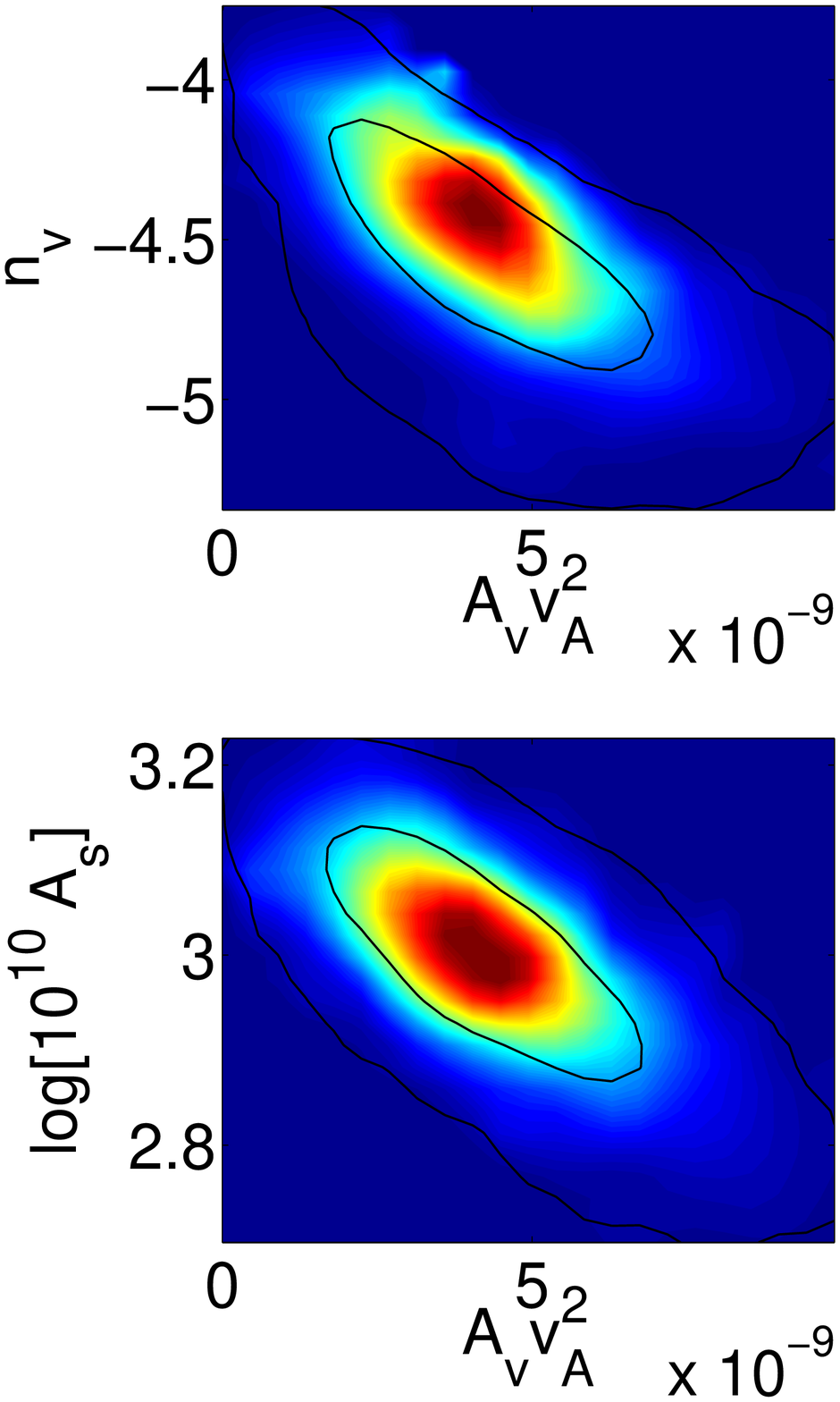}
\centering\includegraphics[scale=.32]{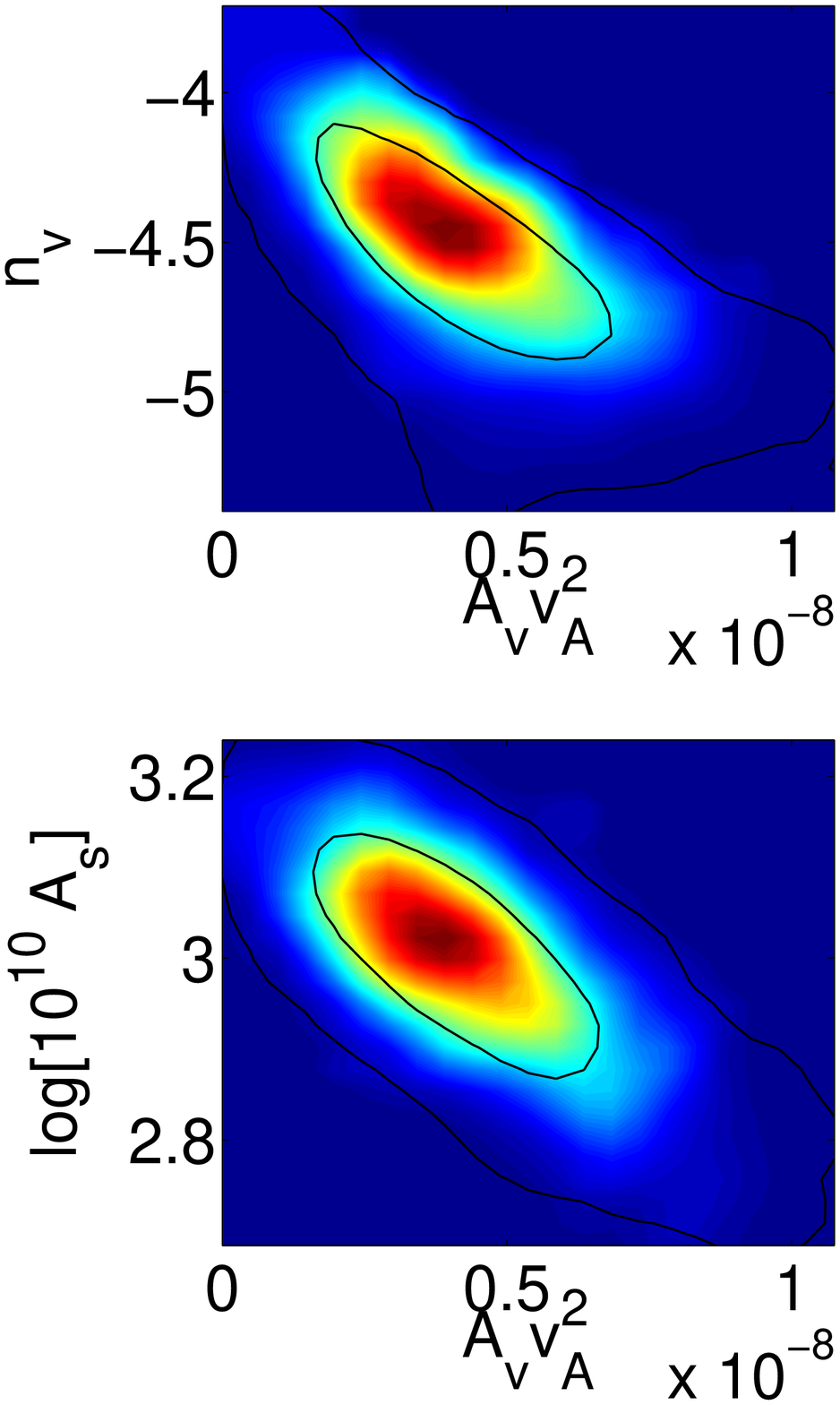}
\caption{Marginalized likelihood in the plane of $A_v\,v^2_A$ vs $\{n_v,\;\log[10^{10}A_s]\,\}$: WILC, HILC and NILC (from left to right), solid curves denote 1$\sigma$ and 2$\sigma$ contours.}
\label{likelihood2}
\end{figure}

\begin{figure}[htb!]
\includegraphics[scale=.18]{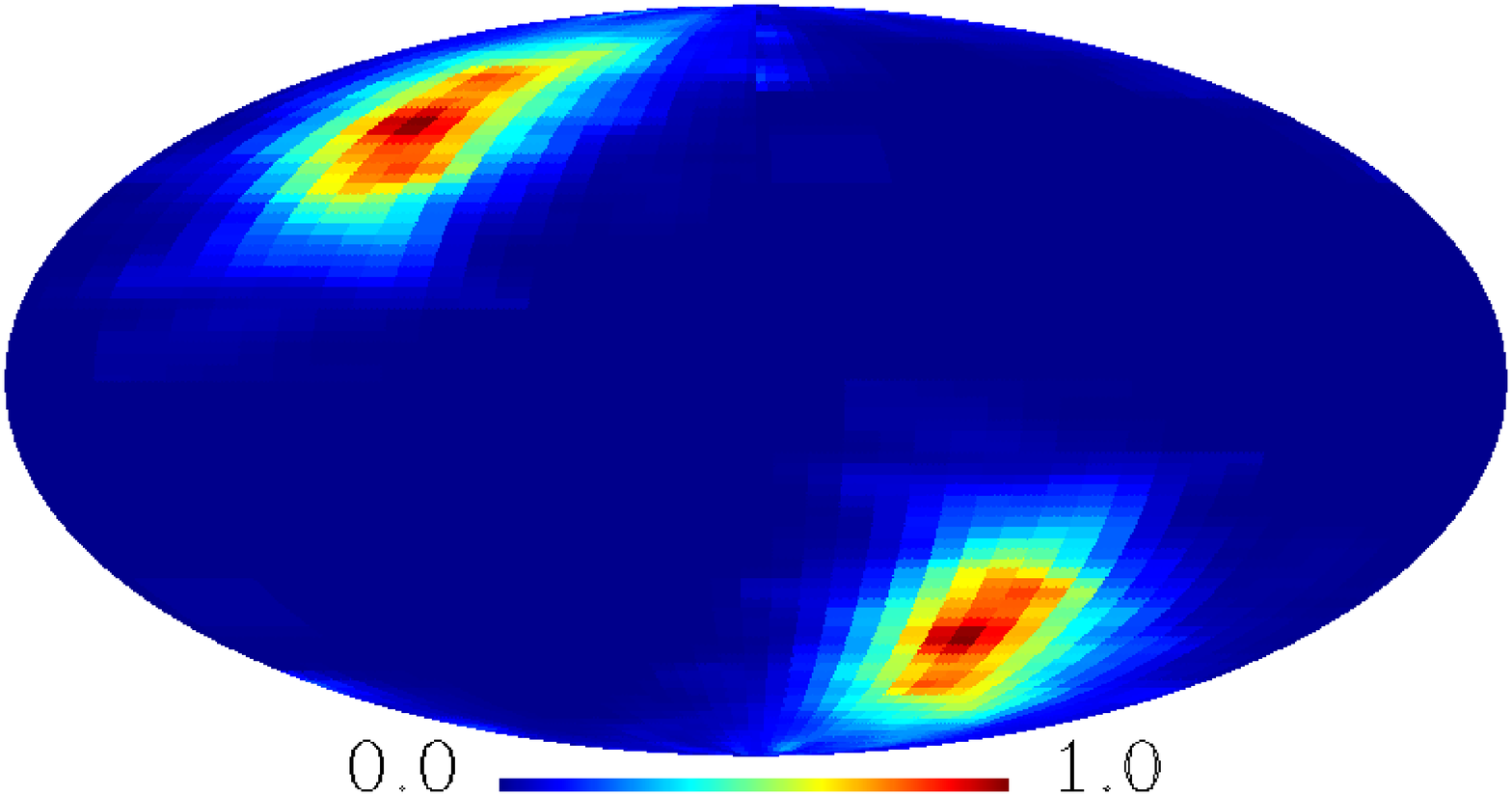}
\includegraphics[scale=.18]{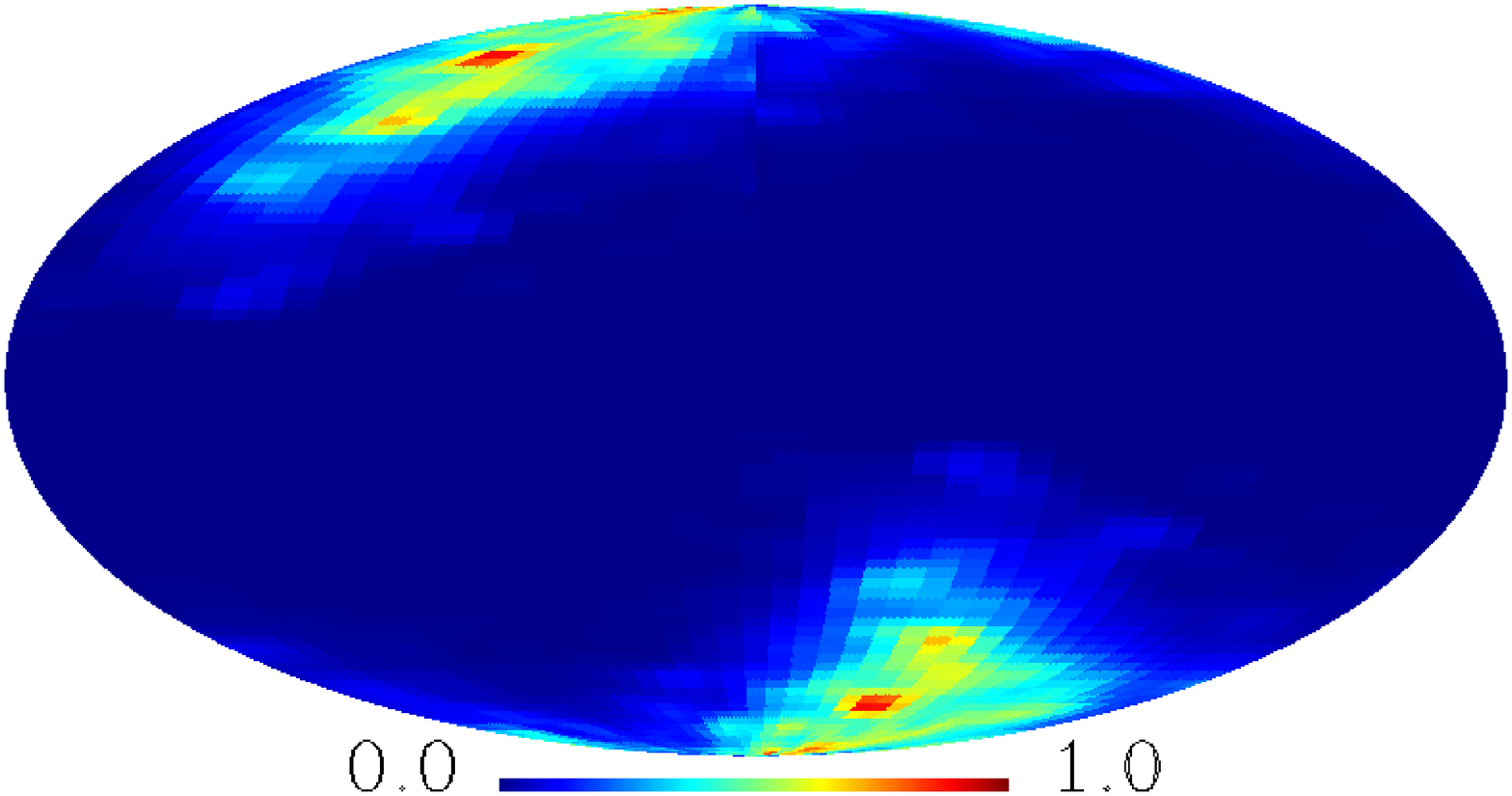}
\includegraphics[scale=.18]{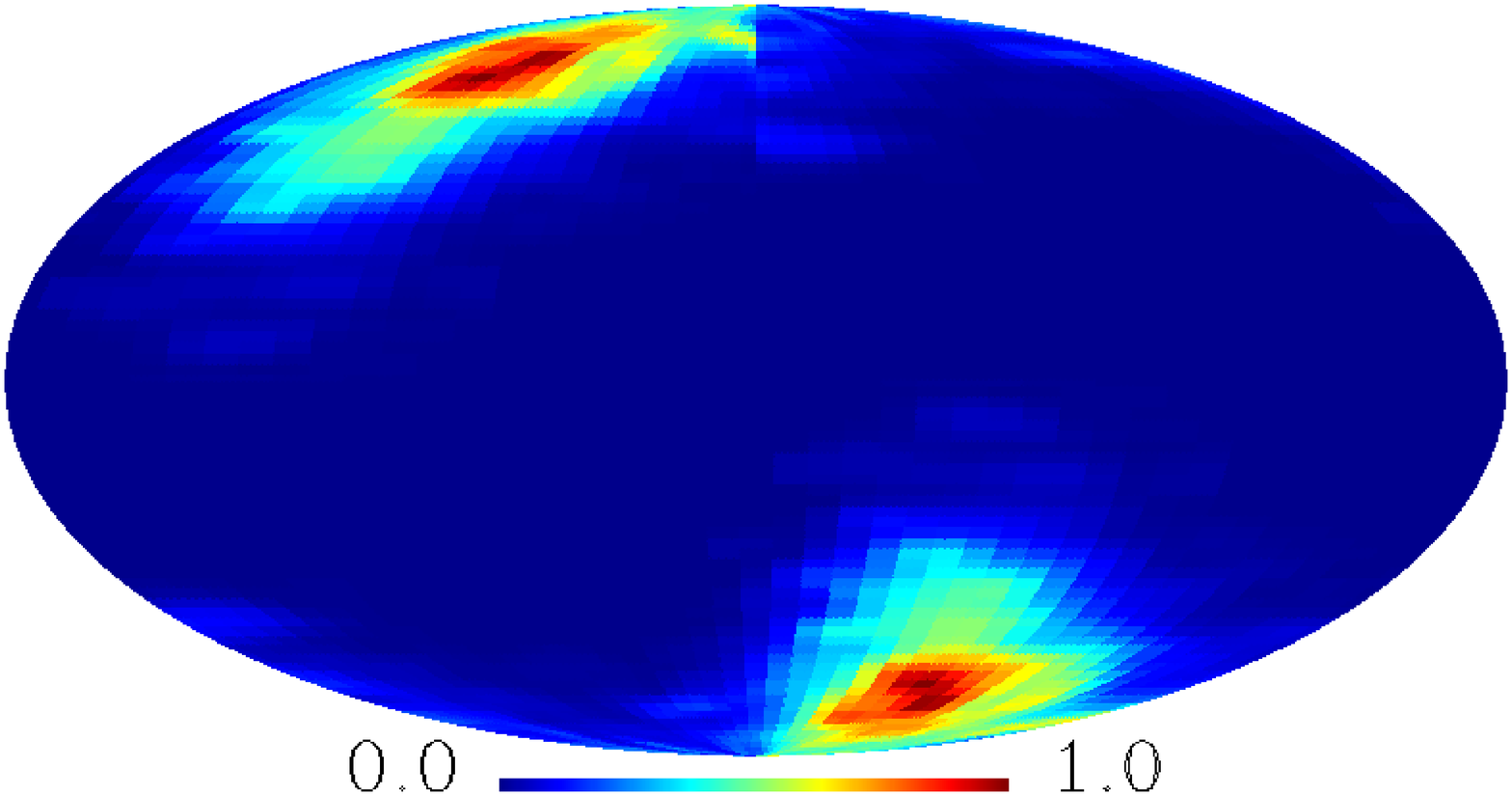}
\caption{Marginalized likelihood of PMF direction: WILC, HILC, and NILC (from left to right), normalized to its peak} 
\label{L_direction}
\end{figure}

By making small modifications to the \texttt{CosmoMC} package \cite{CosmoMC}, we have included the parameter 
$\lambda_\alpha \in \{A_v v^2_A, n_v, \theta_B, \phi_B\}$, and computed the likelihood function given by Eq. \ref{full_likelihood}.
Note that the power spectrum as well as correlations depend on $A_v v^2_A$ and $n_v$ (see Eq. \ref{bar_D0}). For data constraint, we have used the CMB power spectra of the recent CMB observations (WMAP5YR + ACBAR + QUaD) \cite{WMAP5:basic_result,WMAP5:powerspectra,ACBAR,ACBAR2008,QUaD1,QUaD2,QUaD:instrument}, and the correlations estimated from ILC maps.  In Table \ref{parameters}, we summarize 1$\sigma$ constraint and the best-fit values.
As discussed in Sec. \ref{estimators}, our estimators are insensitive to the parity of a PMF direction. 
Hence, we quote two best-fit PMF directions, which are equally likely. 
In Fig. \ref{likelihood1}, we show the marginalized likelihood (solid lines) and mean likelihood (dotted lines) of $\{A_v\,v^2_A, n_v, \theta_B, \phi_B\}$. 
In Fig. \ref{likelihood2}, we show likelihood in the plane of $A_v\,v^2_A$ versus $n_v$, which are highly correlated Alfv\'en wave parameters. We find there exists little correlation between Alfv\'en wave parameters and $\Lambda$CDM parameters, except for the scalar perturbation amplitude $A_s$. In Fig. \ref{likelihood2}, we show marginalized likelihood in the plane of $A_v\,v^2_A$ versus $\log[10^{10} A_s]$. 
Marginalized likelihood of PMF direction is shown in Fig. \ref{L_direction}. 
Note that a pair of directions  associated by $(\theta_B,\phi_B)\leftrightarrow (\pi-\theta_B,\phi_B+\pi)$ possess equal likelihood. 
As shown in Table \ref{parameters} and Fig. \ref{likelihood1}, \ref{likelihood2}, \ref{L_direction},
The result from various ILC maps are similar with each other. Since each ILC map contains distinct residual foregrounds and noise, 
consistency among the analysis with various ILC maps indicates that our results are affected insignificantly by residual foregrounds and noise in ILC maps.

For good exploration on tails of the parameter $A_v v^2_A$, we have also set \texttt{CosmoMC} to check convergence for confidence intervals of $A_v v^2_A$. In Table \ref{3sigma}, we show 3$\sigma$ constraints on $A_v v^2_A$. Using our 3$\sigma$ limit on $A_v\,v^2_A$ and the upper limit on $v_A$ (refer to Sec. 3), we find  $5\times 10^{-4}\lesssim A_v$ (refer to Eq. \ref{P} for the definition of $A_v$). Taking into account Eq. \ref{P}, we find the amplitude of primordial vector perturbation is equal to $A_v k^3_0$.
Hence, we impose the lower bound $4\times 10^{-12}$ on the amplitude of primordial vector perturbation.
\begin{table}[htb!]
\centering\caption{$A_v v^2_A$ estimation at 3$\sigma$ confidence level}
\label{3sigma}
\begin{tabular}{cc}
\hline\hline 
Power Spectra + ILC & 3$\sigma$ confidence interval\\
\hline
WILC  &  $5.8\times 10^{-10}<A_v\,v^2_A<1.01\times 10^{-8}$\\ 
HILC  &  $5.73\times 10^{-10}<A_v\,v^2_A<0.99\times 10^{-8}$\\ 
NILC  &  $5.05\times 10^{-10}<A_v\,v^2_A<1.09\times 10^{-8}$\\ 
\hline
\end{tabular}
\end{table}

If there was a primordial magnetic field at the recombination epoch or after the Universe was re-ionized, it would induce Faraday rotation in CMB polarization \cite{PMF_B,PMF_Faraday_obs}. Forecast on the PLANCK data constraint shows that we will be able to constrain a primordial magnetic field $\sim 2\times 10^{-9}\;\mathrm{Gauss}$ by investigating Faraday rotation effect in polarization data \cite{PMF_Faraday_obs}. 
However, we have not considered Faraday rotation effect in this work, since the signal-to-noise ratio in the currently available CMB polarization data is not high.

\section{Discussion}
\label{Discussion}
We have constrained primordial magnetic field and primordial vector perturbation, by investigating the imprints of cosmological Alfv\'en waves on CMB anisotropy. We find there is little degeneracy between cosmological Alfv\'en wave parameters and $\Lambda$CDM parameters except for a scalar perturbation amplitude $A_s$. 
The results obtained with various ILC maps are consistent with each other. Therefore, we believe our results have not been affected significantly by residual foregrounds or anisotropic noise in ILC maps. Using our result and the upper bound on $v_A$ from the total energy density constraint, we impose a lower bound on the primordial vector perturbation $5\times 10^{-4}\lesssim A_v$ at the pivot scale $k_0=0.002/\mathrm{Mpc}$.

Since Alfv\'en velocity $v_A$ is proportional to the magnitude of a primordial magnetic field, direct constraints on Alfv\'en velocity $v_A$ may be imposed by investigating Faraday rotation effect in CMB polarization data \cite{PMF_B,PMF_Faraday_obs}.
When the data from the PLANCK surveyor \cite{Planck_bluebook} are available, we are going to constrain 
Alfv\'en waves and Faraday rotation simultaneously, and hence impose a stronger constraint on the primordial vector perturbation.

\ack
We thank anonymous referees for helpful comments, which lead to sharpening this work.
We thank T. Kahniashvili and G. Lavrelashvili for useful discussions. 
We also thank J. Delabrouille and his colleagues for making their whole-sky CMB map (NILC) publically available. 
We acknowledge the use of the Legacy Archive for Background Data Analysis (LAMBDA) and
ACBAR, and QUaD data.
Our data analysis made the use of HEALPix \cite{HEALPix:Primer,HEALPix:framework} and the \texttt{CosmoMC} package.
This work is supported by FNU grant 272-06-0417, 272-07-0528 and 21-04-0355. 

\section{Reference}
\bibliographystyle{unsrt}
\bibliography{/home/tac/jkim/Documents/bibliography}

\begin{thebibliography}{10}

\bibitem{Magnetic_Fields_Lyman}
A.~M. {Wolfe}, K.~M. {Lanzetta}, and A.~L. {Oren}.
\newblock {Magnetic fields in damped Ly-alpha systems}.
\newblock {\em \apj}, 388:17--22, March 1992.

\bibitem{Magnetic_Fields_Galaxy}
T.~E. {Clarke}, P.~P. {Kronberg}, and H.~{B{\"o}hringer}.
\newblock {A New Radio-X-Ray Probe of Galaxy Cluster Magnetic Fields}.
\newblock {\em \apjl}, 547:L111--L114, February 2001.

\bibitem{Magnetic_Fields_Origin}
L.~M. {Widrow}.
\newblock {Origin of galactic and extragalactic magnetic fields}.
\newblock {\em Reviews of Modern Physics}, 74:775--823, 2002.

\bibitem{Magnetic_Fields_Faraday}
Y.~{Xu}, P.~P. {Kronberg}, S.~{Habib}, and Q.~W. {Dufton}.
\newblock {A Faraday Rotation Search for Magnetic Fields in Large-scale
  Structure}.
\newblock {\em \apj}, 637:19--26, January 2006.

\bibitem{PMF_Cosmic_transition}
T.~{Vachaspati}.
\newblock {Magnetic fields from cosmological phase transitions.}
\newblock {\em Physics Letters B}, 265:258--261, August 1991.

\bibitem{PMF_Electroweak}
G.~{Baym}, D.~{B{\"o}deker}, and L.~{McLerran}.
\newblock {Magnetic fields produced by phase transition bubbles in the
  electroweak phase transition}.
\newblock {\em \prd}, 53:662--667, January 1996.

\bibitem{PMF_QCD}
J.~M. {Quashnock}, A.~{Loeb}, and D.~N. {Spergel}.
\newblock {Magnetic field generation during the cosmological QCD phase
  transition}.
\newblock {\em \apjl}, 344:L49--L51, September 1989.

\bibitem{PMF_quark}
B.~{Cheng} and A.~V. {Olinto}.
\newblock {Primordial magnetic fields generated in the quark-hadron
  transition}.
\newblock {\em \prd}, 50:2421--2424, August 1994.

\bibitem{cosmic_defect_Vector}
N.~{Turok}, U.-L. {Pen}, and U.~{Seljak}.
\newblock {Scalar, vector, and tensor contributions to CMB anisotropies from
  cosmic defects}.
\newblock {\em \prd}, 58(2):023506--+, July 1998.

\bibitem{cosmic_defect_CMB}
U.~{Seljak}, U.-L. {Pen}, and N.~{Turok}.
\newblock {Polarization of the Microwave Background in Defect Models}.
\newblock {\em \prl}, 79:1615--1618, September 1997.

\bibitem{SPMF_CMB}
D.~{Paoletti}, F.~{Finelli}, and F.~{Paci}.
\newblock {The full contribution of a stochastic background of magnetic fields
  to CMB anisotropies}.
\newblock {\em MNRAS in press}, November 2008.

\bibitem{Foundations_Cosmology}
Viatcheslav Mukhanov.
\newblock {\em Physical Foundations of Cosmology}.
\newblock Cambridge University Press, 1st edition, 2005.

\bibitem{Cosmology}
Steven Weinberg.
\newblock {\em Cosmology}.
\newblock Oxford University Press, 1st edition, 2008.

\bibitem{PMF_acoustic_CMB}
J.~{Adams}, U.~H. {Danielsson}, D.~{Grasso}, and H.~{Rubinstein}.
\newblock {Distortion of the acoustic peaks in the CMBR due to a primordial
  magnetic field}.
\newblock {\em Physics Letters B}, 388:253--258, February 1996.

\bibitem{inhomogeneous_Alfven}
K.~{Subramanian} and J.~D. {Barrow}.
\newblock {Microwave Background Signals from Tangled Magnetic Fields}.
\newblock {\em Physical Review Letters}, 81:3575--3578, October 1998.

\bibitem{pmf_small}
K.~{Subramanian} and J.~D. {Barrow}.
\newblock {Small-scale microwave background anisotropies arising from tangled
  primordial magnetic fields}.
\newblock {\em \mnras}, 335:L57--L61, September 2002.

\bibitem{pmf_pol}
K.~{Subramanian}, T.~R. {Seshadri}, and J.~D. {Barrow}.
\newblock {Small-scale cosmic microwave background polarization anisotropies
  due to tangled primordial magnetic fields}.
\newblock {\em \mnras}, 344:L31--L35, September 2003.

\bibitem{DKY}
R.~{Durrer}, T.~{Kahniashvili}, and A.~{Yates}.
\newblock Microwave background anisotropies from {A}lfv{\'e}n waves.
\newblock {\em \prd}, 58:123004, 1998.

\bibitem{Alfven_nl}
K.~{Subramanian} and J.~D. {Barrow}.
\newblock {Magnetohydrodynamics in the early universe and the damping of
  nonlinear Alfv{\'e}n waves}.
\newblock {\em \prd}, 58(8):083502--+, October 1998.

\bibitem{KLR}
T.~{Kahniashvili}, G.~{Lavrelashvili}, and B.~{Ratra}.
\newblock {CMB temperature anisotropy from broken spatial isotropy due to a
  homogeneous cosmological magnetic field}.
\newblock {\em \prd}, 78(6):063012, September 2008.

\bibitem{WMAP5:basic_result}
G.~Hinshaw and et~al.
\newblock Five-year wilkinson microwave anisotropy probe ({WMAP}) observations:
  Data processing, sky maps, and basic results.
\newblock 2008.
\newblock arXiv:0803.0732.

\bibitem{WMAP5:powerspectra}
M.~R. Nolta and et~al.
\newblock Five-year {W}ilkinson {M}icrowave {A}nisotropy probe ({WMAP})
  observations: Angular power spectra.
\newblock {\em submitted to {\apjs}}, 2008.
\newblock arXiv:0803.0593.

\bibitem{WMAP5:parameter}
J.~{Dunkley}, E.~{Komatsu}, M.~R. {Nolta}, D.~N. {Spergel}, D.~{Larson},
  G.~{Hinshaw}, L.~{Page}, C.~L. {Bennett}, B.~{Gold}, N.~{Jarosik}, J.~L.
  {Weiland}, M.~{Halpern}, R.~S. {Hill}, A.~{Kogut}, M.~{Limon}, S.~S. {Meyer},
  G.~S. {Tucker}, E.~{Wollack}, and E.~L. {Wright}.
\newblock {Five-Year Wilkinson Microwave Anisotropy Probe Observations:
  Likelihoods and Parameters from the WMAP Data}.
\newblock {\em \apjs}, 180:306--329, February 2009.

\bibitem{ACBAR}
M.~C. Runyan, P.~A.~R. Ade, R.~S. Bhatia, J.~J. Bock, M.~D. Daub, J.~H.
  Goldstein, C.~V. Haynes, W.~L. Holzapfel, C.~L. Kuo, A.~E. Lange, J.~Leong,
  M.~Lueker, M.~Newcomb, J.~B. Peterson, J.~Ruhl, G.~Sirbi, E.~Torbet,
  C.~Tucker, A.~D. Turner, and D.~Woolsey.
\newblock Acbar: The arcminute cosmology bolometer array receiver.
\newblock {\em \apjs}, 149:265, 2003.

\bibitem{ACBAR2008}
C.~L. Reichardt, P.~A.~R. Ade, J.~J. Bock, J.~R. Bond, J.~A. Brevik, C.~R.
  Contaldi, M.~D. Daub, J.~T. Dempsey, J.~H. Goldstein, W.~L. Holzapfel, C.~L.
  Kuo, A.~E. Lange, M.~Lueker, M.~Newcomb, J.~B. Peterson, J.~Ruhl, M.~C.
  Runyan, and Z.~Staniszewski.
\newblock High resolution cmb power spectrum from the complete {ACBAR} data
  set.
\newblock {\em To be submitted to {\apj}}, 2008.
\newblock arXiv:0801.1491.

\bibitem{QUaD1}
P.~Ade, J.~Bock, M.~Bowden, M.~L. Brown, G.~Cahill, J.~E. Carlstrom, P.~G.
  Castro, S.~Church, T.~Culverhouse, R.~Friedman, K.~Ganga, W.~K. Gear,
  J.~Hinderks, J.~Kovac, A.~E. Lange, E.~Leitch, S.~J. Melhuish, J.~A. Murphy,
  A.~Orlando, R.~Schwarz, C.~O'Sullivan, L.~Piccirillo, C.~Pryke, N.~Rajguru,
  B.~Rusholme, A.~N. Taylor, K.~L. Thompson, E.~Y.~S. Wu, and M.~Zemcov.
\newblock First season quad cmb temperature and polarization power spectra.
\newblock {\em {\apj}}, 674:22, 2008.

\bibitem{QUaD2}
{QUaD collaboration: C.~Pryke}, P.~{Ade}, J.~{Bock}, M.~{Bowden}, M.~L.
  {Brown}, G.~{Cahill}, P.~G. {Castro}, S.~{Church}, T.~{Culverhouse},
  R.~{Friedman}, K.~{Ganga}, W.~K. {Gear}, S.~{Gupta}, J.~{Hinderks},
  J.~{Kovac}, A.~E. {Lange}, E.~{Leitch}, S.~J. {Melhuish}, Y.~{Memari}, J.~A.
  {Murphy}, A.~{Orlando}, R.~{Schwarz}, C.~{O'Sullivan}, L.~{Piccirillo},
  N.~{Rajguru}, B.~{Rusholme}, A.~N. {Taylor}, K.~L. {Thompson}, A.~H.
  {Turner}, E.~Y.~S. {Wu}, and M.~{Zemcov}.
\newblock {Second and third season QUaD CMB temperature and polarization power
  spectra}.
\newblock {\em ArXiv e-prints}, May 2008.

\bibitem{QUaD:instrument}
{QUaD collaboration: J.~Hinderks}, P.~{Ade}, J.~{Bock}, M.~{Bowden}, M.~L.
  {Brown}, G.~{Cahill}, J.~E. {Carlstrom}, P.~G. {Castro}, S.~{Church},
  T.~{Culverhouse}, R.~{Friedman}, K.~{Ganga}, W.~K. {Gear}, S.~{Gupta},
  J.~{Harris}, V.~{Haynes}, J.~{Kovac}, E.~{Kirby}, A.~E. {Lange}, E.~{Leitch},
  O.~E. {Mallie}, S.~{Melhuish}, A.~{Murphy}, A.~{Orlando}, R.~{Schwarz},
  C.~{O' Sullivan}, L.~{Piccirillo}, C.~{Pryke}, N.~{Rajguru}, B.~{Rusholme},
  A.~N. {Taylor}, K.~L. {Thompson}, C.~{Tucker}, E.~Y.~S. {Wu}, and
  M.~{Zemcov}.
\newblock Quad: A high-resolution cosmic microwave background polarimeter.
\newblock {\em ArXiv e-prints}, May 2008.

\bibitem{Chen}
G.~{Chen}, P.~{Mukherjee}, T.~{Kahniashvili}, B.~{Ratra}, and Y.~{Wang}.
\newblock {Looking for Cosmological Alfv{\'e}n Waves in Wilkinson Microwave
  Anisotropy Probe Data}.
\newblock {\em {\apj}}, 611:655--659, August 2004.

\bibitem{Naselsky:PMF}
P.~D. {Naselsky}, L.-Y. {Chiang}, P.~{Olesen}, and O.~V. {Verkhodanov}.
\newblock {Primordial Magnetic Field and Non-Gaussianity of the One-Year
  Wilkinson Microwave Anisotropy Probe Data}.
\newblock {\em \apj}, 615:45--54, November 2004.

\bibitem{PMF_anomaly}
A.~{Bernui} and W.~S. {Hip{\'o}lito-Ricaldi}.
\newblock {Can a primordial magnetic field originate large-scale anomalies in
  WMAP data?}
\newblock {\em \mnras}, 389:1453--1460, September 2008.

\bibitem{Spacetime}
Sean Carroll.
\newblock {\em Spacetime and Geometry: An Introduction to General Relativity}.
\newblock Benjamin Cummings, 2003.

\bibitem{Fixen:dipole}
D.~J. Fixsen, E.~S. Cheng, J.~M. Gales, J.~C. Mather, R.~A. Shafer, and E.~L.
  Wright.
\newblock The cosmic microwave background spectrum from the full {COBE} {FIRAS}
  data set.
\newblock {\em \apj}, 473:576, 1996.

\bibitem{WMAP3:parameter}
D.~N. Spergel and et~al.
\newblock Wilkinson {M}icrowave {A}nisotropy probe {WMAP} three year results:
  Implications for cosmology.
\newblock {\em \apjs}, 170:377, 2007.

\bibitem{PMF_Constraint}
J.~D. {Barrow}, P.~G. {Ferreira}, and J.~{Silk}.
\newblock {Constraints on a Primordial Magnetic Field}.
\newblock {\em Physical Review Letters}, 78:3610--3613, May 1997.

\bibitem{Math_methods}
S.~J. Bence~(Author) K.~F. Riley M. P. Hobson~(Author).
\newblock {\em Mathematical Methods for Physics and Engineering: A
  Comprehensive Guide}.
\newblock Cambridge University Press, 3rd edition, 2006.

\bibitem{WMAP1:parameter_method}
L.~Verde, H.~V. Peiris, D.~N. Spergeland M. Noltaand C. L. Bennettand M.
  Halpernand G. Hinshawand N. Jarosikand A. Kogutand~M. Limon, and S.~S.
  Meyerand L. Pageand G. S. Tuckerand E. Wollackand E.~L. Wright.
\newblock First year wilkinson microwave anisotropy probe ({WMAP})
  observations: Parameter estimation methodology.
\newblock {\em \apjs}, 148:195, 2003.

\bibitem{CosmoMC}
Antony Lewis and Sarah Bridle.
\newblock Cosmological parameters from {CMB} and other data: a {Monte-Carlo}
  approach.
\newblock {\em \prd}, 66:103511, 2002.

\bibitem{WMAP3:temperature}
G.~Hinshaw and et~al.
\newblock Three-year {Wilkinson Microwave Anisotropy Probe} ({WMAP})
  observations: Temperature analysis.
\newblock {\em \apjs}, 170:288, 2007.

\bibitem{WMAP5:foreground}
B.~Gold and et~al.
\newblock Five-year wilkinson microwave anisotropy probe ({WMAP}) observations:
  Galactic foreground emission.
\newblock {\em submitted to {\apjs}}, 2008.
\newblock arXiv:0803.0715.

\bibitem{HILCT}
J.~{Kim}, P.~{Naselsky}, and P.~R. {Christensen}.
\newblock {CMB map derived from the WMAP data through harmonic internal linear
  combination}.
\newblock {\em \prd}, 77(10):103002--+, May 2008.

\bibitem{HILCP}
J.~{Kim}, P.~{Naselsky}, and P.~R. {Christensen}.
\newblock {CMB polarization map derived from the WMAP 5year data through the
  harmonic internal linear combination method}.
\newblock {\em \prd}, 79(2):023003--+, January 2009.
\newblock http://www.nbi.dk/$\sim$jkim/hilc.

\bibitem{NILC}
J.~Delabrouille, J.-F. Cardoso, M.~Le Jeune, M.~Betoule, G.~Fay, and
  F.~Guilloux.
\newblock A full sky, low foreground, high resolution {CMB} map from {WMAP}.
\newblock {\em submitted to AA}, 2008.
\newblock arXiv:0807.0773.

\bibitem{PMF_B}
C.~{Sc{\'o}ccola}, D.~{Harari}, and S.~{Mollerach}.
\newblock {B polarization of the CMB from Faraday rotation}.
\newblock {\em \prd}, 70(6):063003--+, September 2004.

\bibitem{PMF_Faraday_obs}
J.~R. {Kristiansen} and P.~G. {Ferreira}.
\newblock {Constraining primordial magnetic fields with CMB polarization
  experiments}.
\newblock {\em \prd}, 77(12):123004--+, June 2008.

\bibitem{Planck_bluebook}
European~Space Agency.
\newblock Planck: The scientific programme (blue book).
\newblock
  http://www.rssd.esa.int/SA/PLANCK/docs/Bluebook-ESA-SCI
  2005.
\newblock Version 2.

\bibitem{HEALPix:Primer}
K.~M. Gorski, B.~D. Wandelt, F.~K. Hansen, E.~Hivon, and A.~J. Banday.
\newblock The {HEALPix} primer.
\newblock astro-ph/9905275, 1999.

\bibitem{HEALPix:framework}
K.~M. Gorski, E.~Hivon, A.~J. Banday, B.~D. Wandelt, F.~K. Hansen, M.~Reinecke,
  and M.~Bartelman.
\newblock {HEALPix} -- a framework for high resolution discretization, and fast
  analysis of data distributed on the sphere.
\newblock {\em \apj}, 622:759, 2005.
\newblock http://healpix.jpl.nasa.gov.

\end{thebibliography}

\begin{appendix}
\section{incomplete sky coverage}
\label{leakage}
Incomplete sky coverage produces fictitious correlations by destroying the orthogonality of spherical harmonics.
Spherical harmonic coefficients from partial sky coverage is related to the true ones as follows:
$a_{lm} =\sum W_{ll'mm'} a^{\mathrm{true}}_{l'm'},$
where 
$ W_{ll'mm'}=\int d^2\hat{\mathbf{n}}\,W(\hat{\mathbf{n}}) Y^*_{lm}(\hat{\mathbf{n}})\,Y_{l'm'}(\hat{\mathbf{n}})$, and 
$W(\hat{\mathbf{n}})$ is zero in the masked region and one elsewhere.
To estimate the amount of leakage, we assume CMB is purely Gaussian and has zero correlation:
\begin{eqnarray*} 
\langle (a^{\mathrm{true}}_{lm})^* a^{\mathrm{true}}_{l',m'}\rangle = C_l\,\delta_{ll'} \delta_{mm'}.
\end{eqnarray*}
We find the expectation value of $D^3_l$ of the pure Gaussian CMB is
\begin{eqnarray} 
\lefteqn{\langle D^3_l\rangle=\frac{1}{2l+1}\sum_m\langle a^*_{l,m} a_{l+2,m}\rangle}\label{D_leakage}\\
&=&\frac{1}{2l+1}\sum_m\sum_{l'm'} C_{l'} W^*_{ll'mm'} W_{l+2,l'mm'}\nonumber\\
&\approx& \frac{C_{l+1} }{2l+1}\sum_{m}\sum_{l'm'}\int d^2\hat{\mathbf{n}}\,W^*(\hat{\mathbf{n}}) Y_{lm}(\hat{\mathbf{n}})\,Y^*_{l'm'}(\hat{\mathbf{n}})\int d^2\mathbf{\hat{n}'}\,W(\mathbf{\hat{n}'}) Y^*_{l+2\,m}(\mathbf{\hat{n}'})\,Y_{l'm'}({\mathbf{\hat{n}'}})\nonumber\\
&=&\frac{C_{l+1} }{2l+1} \sum_{m}\int d^2\hat{\mathbf{n}}\,W^*(\hat{\mathbf{n}})Y_{lm}(\hat{\mathbf{n}})\int d^2\mathbf{\hat{n}'} W(\mathbf{\hat{n}'}) \,Y^*_{l+2\,m}(\mathbf{\hat{n}'})\delta(\hat{\mathbf{n}}-\mathbf{\hat{n}'})\nonumber\\
&=&\frac{C_{l+1} }{2l+1} \sum_{m} \int d^2\hat{\mathbf{n}}\,|W(\hat{\mathbf{n}})|^2 Y_{lm}(\hat{\mathbf{n}}) Y^*_{l+2\,m}(\hat{\mathbf{n}}).\nonumber
\end{eqnarray}
In the third equality, we have taken $C_{l'}$ out of the summation and equate it to $C_{l+1}$, since $W^*_{ll'mm'} W_{l+2,l'mm'}$ peaks sharply around $l'=l+1$, while $C_{l'}$ varies much slowly in comparison to $W^*_{ll'mm'} W_{l+2,l'mm'}$.
In the fourth equality, we have used the identity $\sum_{l'm'} Y_{l'm'}(\hat{\mathbf{n}}) Y_{l'm'}(\mathbf{\hat{n}'}) =\delta(\hat{\mathbf{n}}-\mathbf{\hat{n}'})$. From Eq. \ref{D_leakage}, we see the leakage from Gaussian CMB to $D^3_l$ can be estimated by:
\begin{eqnarray*} 
\frac{1}{2l+1}\sum_{m}\int d^2\hat{\mathbf{n}}\,|W(\hat{\mathbf{n}})|^2 Y_{lm}(\hat{\mathbf{n}}) Y^*_{l+2,m}(\hat{\mathbf{n}}).\label{Cl_leakage}
\end{eqnarray*}
The value as high as unity indicates that most of Gaussian CMB power is leaked into the fictitious correlation, while a zero value indicates no leakage. 
\begin{table}[htb!]
\centering
\caption{D3 Leakage}\label{D3_leakage}
\begin{tabular}{c|cccc}
\hline
\hline 
$l$ & 10& 20 &100 & 200\\ 
\hline
KQ75&\; $0.18$ &\; $0.18$ &\;$0.17$ &\; $0.17$\\ 
KQ85&\; $0.12$ &\; $0.12$ &\;$0.12$ &\; $0.12$\\
\hline 
\end{tabular}
\end{table}
\end{appendix}
We have estimated $D^i_l$ leakages for the WMAP team's KQ75 and KQ85 mask and show them in Table \ref{D3_leakage}.
We find $D^3_l$ leakage is as high as $\sim 20\%$ and other leakgages are $\sim 2\%$, which are all greater than the expected correlations, given Alfv\'en wave parameters in Table \ref{parameters}.  
Therefore, we find CMB maps, which requires the KQ75 or KQ 85 mask, are not suitable for estimation of correlations induced by cosmological Alfv\'en waves.  
\end{document}